\documentclass[12pt]{article}
\usepackage{graphicx}

\begin{document}

\title{IMPROVING SHORT-TERM EOP PREDICTION USING COMBINATION PROCEDURES}
\author{Zinovy Malkin}
\date{Pulkovo Observatory, St. Petersburg, Russia}
\maketitle

\begin{abstract}
A well known problem with EOP prediction is that a prediction
strategy proved to be the best for some testing period and
prediction length may not remain as such for other period of time.
In this paper we consider possible strategies to combine
EOP predictions made using different analysis technique to obtain
a final prediction with the best accuracy corresponding to the least
prediction error between input predictions.
This approach can be used to improve the short-term real-time EOP forecast.
\end{abstract}

\vfill
\noindent \hrule width 0.4\textwidth
~\vskip 0.2ex
\noindent {\small Presented at the Journ\'ees 2008: Astrometry, Geodynamics and Astronomical  Reference Systems, Dresden, Germany, 22-24 Sep 2008}
\eject

\section{INTRODUCTION}

\begin{figure}[!ht]
\centering
\includegraphics[clip,width=\hsize]{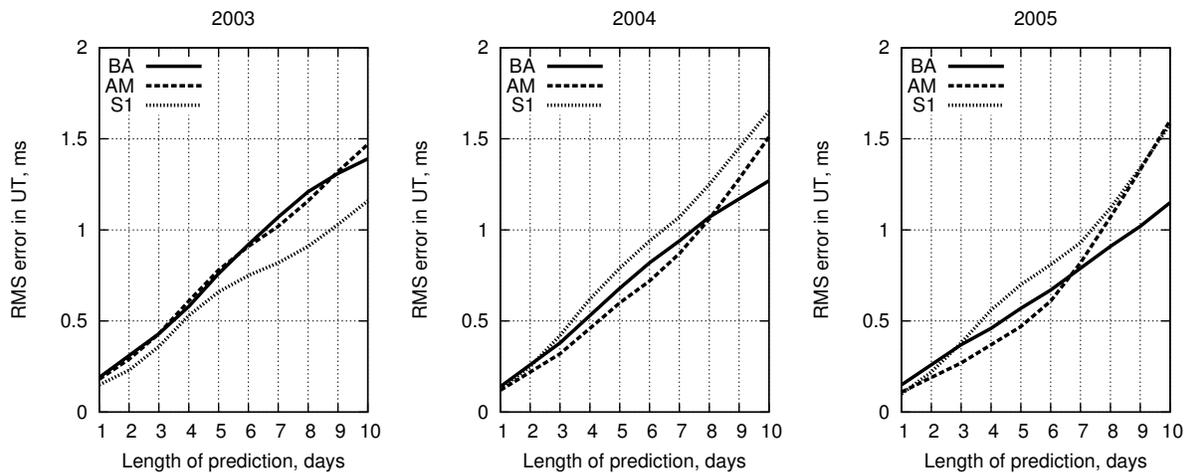}
\caption{Results of actual UT1 predictions for three years.
BA~-- Bulletin A prediction, AM~-- prediction of the same NEOS series made by the author,
S1~-- author's prediction of the SLR series computed at the Institute of Applied Astronomy.
Last two predictions were made making use of method developed by Malkin \& Skurikhina, 1996}
\label{fig:UT1-actual}
\end{figure}

Prediction of the Earth Orientation Parameters (EOP) is a practically
very important and theoretically very interesting task, one of
the main fields of activity of operational EOP services.
Various methods are developed to compute a highly accurate EOP forecast.
However, a usual and well known problem is that different methods show different
accuracy at different time intervals and prediction lengths.
A method which is the best for short-time prediction may not be such
for long-time prediction, and vice versa.
On the other hand, a method which was proven to be the best for a testing
period of time may not remain as such for the coming period.
As an example, results of actual UT1 predictions are shown in Figure~\ref{fig:UT1-actual}.
One can see that different predictions show the best results for different
years and length of prediction.

A practical consequence is that a prediction procedure optimally adjusted
for some time period may not remain the best for the following time period.
Evidently, the main reason of this is that the Earth rotation is more
complicated process than we are able describe by our forecast models.
So, the question is whether we can improve our prediction strategy to make
it more robust to unpredictable behavior of the Earth?

In this paper we try to use combining procedures to achieve this goal.
We will concentrate on short-term prediction up to 5-day length as evidently
the most interesting for practical needs, GNSS applications in the first place.

\section{TEST PREDICTIONS}

For testing purposes, nine series of everyday predictions for the period
from 1 October 2006 through 31 December 2007 making use of prediction strategy
developed at the Institute of Applied Astronomy (Malkin \& Skurikhina, 1996).
This method includes three prediction techniques:
\begin{enumerate}
\item  Least squares fitting for trend and several harmonics,
\item  Autoregression.
\item  Autoregressive Integrated Moving Average.
\end{enumerate}

Final prediction is composed of several segments computed using different methods
and/or model parameters and merged in one continuous series using boundary conditions.
The main model parameters that can be varied to adjust the prediction procedures are the following:
\begin{enumerate}
\item  base interval,
\item  number and period of harmonics,
\item  trend order,
\item  AR order,
\item  IMA order,
\end{enumerate}

Abovementioned nine prediction series were computed using various sets of listed parameters
aside from item 2 which remains the same for all the predictions.

\section{COMBINED PREDICTIONS}

Using prediction time series computed as described above, three combined predictions
were computed for the period from 1 January 2007 through 31 December 2007.
They are the following:
\begin{description}
\item{C1} ---  average of input prediction as proposed and tested by Luzum et al., 2007,
and Schuh et al., 2008.
\item{C2} ---  best previous prediction for given length. Normally, we compute EOP prediction
on the day of the last observed epoch (``today'').
To compute $n$-day ahead prediction we examine the set of predictions made $n$ days ago and
select one that predicts today EOP most accurately.
Then we compute our today $n$-day ahead prediction making use of the method used for computation
of the best prediction made $n$ days ago.
\item{C3} ---  best yesterday prediction.  To compute this prediction we examine the set of predictions
made yesterday and use for today prediction the method which corresponds to the prediction made
yesterday and predicts today EOP most accurately.
\end{description}

\begin{figure}[p]
\centering
\includegraphics[clip,width=\hsize]{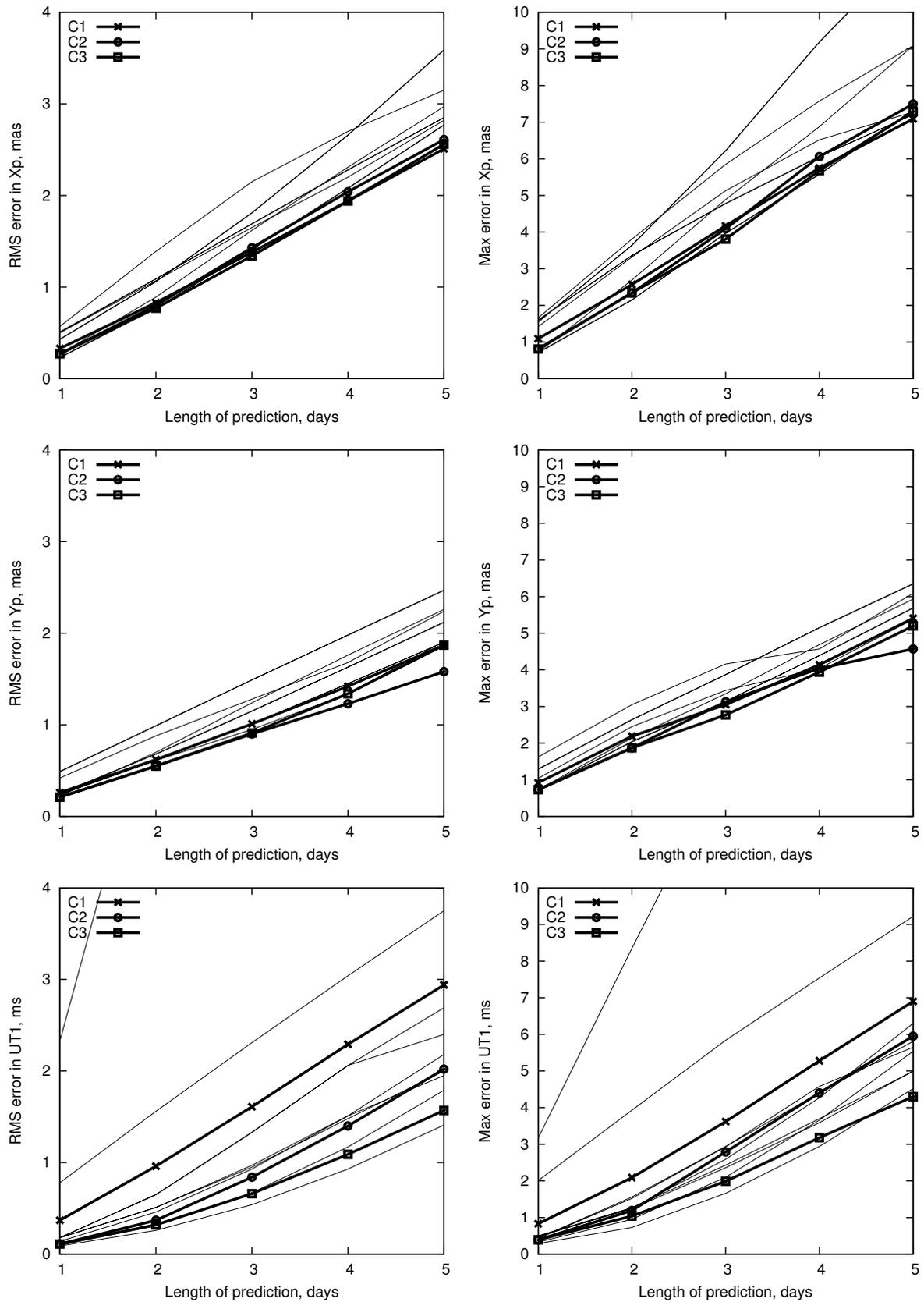}
\caption{The RMS and maximum errors of the input and combined predictions.}
\label{fig:test}
\end{figure}

Note that C2 and C3 predictions also use C1 results along with the input data.
The accuracy of the nine test predictions and three combinations are shown in Fig.~\ref{fig:test}.

From this test one can see that all the tested combined predictions show similar accuracy for
the Pole coordinates, better than accuracy of input series. Marginal advantage of C2 and C3
can bee noticed.

There is more complicated situation with the UT1 prediction.
The average (C1) prediction does not provide a satisfactory result,
probably because of presence of bad predictions
(note the worst UT1 prediction in the upper left part of corresponding plots).
The C3 (best yesterday prediction) is clearly the best for short-time UT1 prediction.

\section{CONCLUSIONS}

From our study presented in this paper we can draw the following conclusions.
\begin{enumerate}
\item A procedure based on computation of multiply predictions with further combination/choice
 of the best one can provide robust accurate  operational EOP forecast.
\item To achieve the best results, input predictions should be computed using the best methods.
\end{enumerate}

\section{REFERENCES}

\leftskip=5mm
\parindent=-5mm

Malkin, Z., Skurikhina, E., 1996, ``On Prediction of EOP'',
Communications of the Institute of Applied Astronomy RAS, No.~93.

Luzum, B., Wooden, W., McCarthy, D., Schuh, H., Kosek,~W., Kalarus,~M., 2007,
``Ensemble Prediction for Earth Orientation Parameters'', Geophysical Research Abstracts, 9, 04315.

Schuh, H., Kosek, W., Kalarus, M., Akyilmaz, O., Gambis,~D., Gross,~R., Jovanovic,~B.,
Kumakshev,~S., Kutterer,~H., Mendes Cerveira,~P.J., Pasynok,~S.,Zotov, L., 2008,
``Earth Orientation Parameters Prediction Comparison Campaign~-- first summary'',
Geophysical Research Abstracts, 10, EGU2008-A-07644.

\end{document}